\documentclass[12pt]{article}
\usepackage{fullpage,epsfig,graphics,amsbsy,amssymb,cancel,bbm}
\usepackage{psfrag,hyperref,color,slashed}
\usepackage{graphicx}
\usepackage{amsfonts,mathdots,dsfont}
\usepackage{amssymb,amsmath,amscd,mathrsfs}
\usepackage{tikz}
\usetikzlibrary{arrows,chains,matrix,positioning,scopes,snakes}

\makeatletter
\tikzset{join/.code=\tikzset{after node path={%
\ifx\tikzchainprevious\pgfutil@empty\else(\tikzchainprevious)%
edge[every join]#1(\tikzchaincurrent)\fi}}}
\makeatother

\tikzset{>=stealth',every on chain/.append style={join},
         every join/.style={->}}
\tikzset{
    >=stealth',
    punkt/.style={
           rectangle,
           rounded corners,
           draw=black, very thick,
           text width=6.5em,
           minimum height=2em,
           text centered},
    pil/.style={
           ->,
           thick,
           shorten <=2pt,
           shorten >=2pt,}
}

\newcommand{\bea}{\begin{eqnarray}}
\newcommand{\eea}{\end{eqnarray}}
\newcommand{\be}{\begin{eqnarray}}
\newcommand{\ee}{\end{eqnarray}}

\newcommand{\Tr}{\textrm{Tr}}

\DeclareMathAlphabet{\mathpzc}{OT1}{pzc}{m}{it}

\newcommand{\qed}{\nobreak \ifvmode \relax \else
      \ifdim\lastskip<1.5em \hskip-\lastskip
      \hskip1.5em plus0em minus0.5em \fi \nobreak
      \vrule height0.5em width0.5em depth0.00em\fi}

\begin{document}
\thispagestyle{empty}
\begin{flushright} \small
UUITP-19/15
 \end{flushright}
\smallskip
\begin{center} \LARGE
{\bf  $q$-Virasoro constraints in matrix models}
 \\[12mm] \normalsize
{\bf  Anton Nedelin$^{a,b}$ and Maxim Zabzine$^b$} \\[8mm]
 {\small\it
${}^a$Dipartimento di Fisica, Universit\`a di Milano--Bicocca, \\
Piazza della Scienza 3, I-20126 Milano, Italy \\and \\ INFN, sezione di Milano--Bicocca, I-20126 Milano, Italy\\
      \vspace{.5cm}
${}^b$Department of Physics and Astronomy,
     Uppsala university,\\
     Box 516,
     SE-75120 Uppsala,
     Sweden\\
   }
\end{center}
\vspace{7mm}
\begin{abstract}
 \noindent
    The Virasoro constraints play the important
     role in the study of matrix models and in understanding of the relation between matrix models and  CFTs. 
     Recently the localization calculations in supersymmetric gauge theories produced 
      new families of matrix models and we have very limited knowledge about these matrix models. 
        We concentrate on elliptic generalization of hermitian matrix model which corresponds to calculation of partition function on $S^3 \times S^1$ for vector multiplet.
     We derive the $q$-Virasoro constraints for this matrix model.  We also observe some interesting algebraic properties of the $q$-Virasoro algebra. 
\end{abstract}

\eject
\normalsize

\section{Introduction}\label{sec_intro}

The matrix models can be thought of  as gauge theories in zero dimensions. 
The most well-know example is given by the hermitian matrix model. We
 define the generating function for this model as an integral over hermitian $N\times N$ matrices $M$
 \be
 Z_N^{\rm herm}(\{t\}) = \int dM ~ e^{\sum\limits_{s=0}^\infty \frac{t_s}{s!} \Tr(M^s)}~,\label{herm-mat-mat}
\ee
where the measure $dM$ is fixed in such way that it is invariant under the symmetry $M\rightarrow U M U^\dagger$,
where $U$ is a $U(N)$-matrix. Alternatively  the matrix model (\ref{herm-mat-mat}) can be rewritten as the integral
over eigenvalues $\lambda_i$
\be
 Z_N^{\rm herm}(\{t\})= \int \prod\limits_{i=1}^N d\lambda_i ~\prod_{i < j} (\lambda_i - \lambda_j)^2 ~e^{\sum\limits_{s=0}^\infty \frac{t_s}{s!} \sum\limits_i \lambda_i^s}~,\label{hem-mat-eig}
\ee
 where the measure of integration is given by the well-known Vandermonde determinant.  This matrix model is  
  well studied and one of the central properties of $Z_N^{\rm herm}(\{t\})$
  is that it is annihilated by an infinite set of differential operator in $t_s$'s which are known
  as Virasoro constraints \cite{Dijkgraaf:1990rs,Mironov:1990im}. 

The natural trigonometric generalization of the hermitian matrix model is defined by the following integral
\be
 Z_N^{\rm trig} (\{ t \})= \int \prod\limits_{i=1}^N d\lambda_i ~\prod_{i < j} \sinh^2 (\lambda_i - \lambda_j) ~e^{\sum\limits_{s=0}^\infty \frac{t_s}{s!} \sum\limits_i \lambda_i^s}~. 
\ee
 If we set all the $t_s$'s to zero except for $s=2$ then this matrix model describes the partition function 
 for $U(N)$ Chern-Simons theory on $S^3$  \cite{Marino:2002fk, Aganagic:2002wv}.  However we may study
 the expectation values of Wilson loops (for a simple knot along the $S^1$-fiber) 
 in different representations and thus $Z_N^{\rm trig} (\{ t \})$ can be regarded as the 
 generating function for the expectation values of Wilson loops in different representations 
(for this simple concrete knot) in Chern-Simons theory.  This matrix  model can be 
also derived through the localization technique applied to supersymmetric Chern-Simons theory 
(supersymmetric vector multiplet in 3D) 
    \cite{Kapustin:2009kz}. 

The elliptic generalization of hermitian matrix model (\ref{hem-mat-eig}) is naturally given by the following formula
\be
 Z_N^{\rm ell}(\{ t\})= \oint \prod\limits_{i=1}^N \frac{d z_i}{z_i} ~\prod_{i < j}  \theta \left ( \frac{z_i}{z_j}; q\right ) \theta \left (\frac{z_j}{z_i}; q \right  )  ~e^{\sum\limits_{s=0}^\infty \frac{t_s}{s!} \sum\limits_i z_i^s}~,\label{ell-matrix-mod}
\ee
where the $\theta$ functions is defined as follows
\be
 \theta (z; q) = \prod_{k=0}^\infty (1- z q^k) (1- z^{-1} q^{k+1})~.\label{def-theta}
\ee
 If we set all $t_n$'s to zero the model (\ref{ell-matrix-mod}) corresponds to the partition function
 of a vector multiplet on $S^3 \times S^1$ \cite{Gadde:2011ia,Dolan:2008qi}. 
  If we allow the supersymmetric Wilson loops then $Z_N^{\rm ell}(\{ t\})$
  can be thought of as the generating function for the expectation values 
  of Wilson loops in different representations.
  Alternatively the matrix model (\ref{ell-matrix-mod}) can be written in 
  other coordinates $z_i = e^{i\lambda_i}$. However the form (\ref{ell-matrix-mod})
  is more standard for the discussion of partition functions  and the
   supersymmetric indices on $S^3 \times S^1$.
    
The hermitian model $Z_N^{\rm herm}(\{t\})$ satisfies the Virasoro constraints,  
see \cite{Morozov:1994hh} for the review of the subject. The natural question 
is whether the trigonometric  $Z_N^{\rm trig} (\{ t \})$
 and elliptic $Z_N^{\rm ell}(\{ t\})$ generalisations  also satisfy some type of Virasoro constraints. The goal of this paper is to answer this question. We will show that $Z_N^{\rm trig} (\{ t \})$ satisfies 
  the Virasoro constraints while the elliptic model $Z_N^{\rm ell}(\{ t\})$ satisfies the deformed $q$-Virasoro constraints. On the way we observe some interesting properties of the $q$-Virasoro algebra. 
 
   The paper is organised as follows:  In section \ref{sec-Vir} we review the derivation of the Virasoro 
   constraints  for the hermitian matrix model and we derive 
    the Virasoro constraints for the trigonometric generalization.  In section \ref{sec-toy}
    we discuss a different approach to the derivation of Virasoro constraints, 
     we introduce the basics of $q$-calculus and derive the $q$-Virasoro constraints for a toy model.
   In  section \ref{sec-q-Vir} we apply these ideas to the elliptic generalization of the 
   hermitian matrix model and derive $q$-Virasoro constraints. Section \ref{s-calculation} 
   contains the technical details of the derivation of the $q$-Virasoro algebra and the discussion 
    of subtleties.  We conclude in section \ref{sec-end} and make some general remarks about our results. 

 \section{Virasoro constraints for hermitian matrix model}\label{sec-Vir}
 
 In this section we review the derivation of the Virasoro constraints in the hermitian matrix model. 
 In our presentation we closely follow the original work \cite{Mironov:1990im}. 
 
 The nature of Virasoro constraints comes from the simple observation that the integral does not change under the change of variables. 
 Let us consider matrix integral 
 \bea
 Z_N^{\rm herm}(\{t \}) = \int \prod_{i} d\lambda_i~  \prod_{i\neq j} (\lambda_i - \lambda_j) ~e^{\sum\limits_{s=0}^\infty \frac{t_s}{s!} \sum\limits_{i} \lambda_i^s}\label{herm-matrix}
\label{hermitian:partition}
 \eea
 with the corresponding saddle-point equation 
 \be
 \sum\limits_{s\geq 1}\frac{t_s}{s!}s \lambda_i^{s-1}+2\sum\limits_{j\neq i}\frac{1}{\lambda_i-\lambda_j}=0~.
 \label{hermitian:eom}
 \ee
 Under the shift
  \bea
   \lambda_i~\rightarrow~\lambda_i + \epsilon_n \lambda_i^{n+1}~, ~~~n \geq -1~. 
   \label{shift}
  \eea
  the effective action in (\ref{hermitian:partition}) changes as the following:
  \be
  \delta\left(\sum\limits_{s\geq 0}\frac{t_s}{s!}\sum\limits_i \lambda_i^s+
  \frac{1}{2}\sum\limits_{i\neq j}\log\left(\lambda_i-\lambda_j\right)^2\right)=\epsilon_n \sum\limits_i
  \left(\sum\limits_{s\geq0}\frac{t_s}{s!}s\lambda_i^{s+n}+\sum\limits_{j\neq i}\frac{\lambda_i^{n+1}-\lambda_j^{n+1}}
  {\lambda_i-\lambda_j}\right)=\nonumber\\
  \epsilon_n \sum\limits_i \lambda_i^{n+1}\left(\sum\limits_{s\geq0}\frac{t_s}{s!}s\lambda_i^{s-1}+
  2\sum\limits_{j\neq i}\frac{1}{\lambda_i-\lambda_j}\right)=0~,
  \ee
  where in the last step we have used equations of motion (\ref{hermitian:eom}). Hence (\ref{shift}) is 
  the on-shell symmetry of the partition function (\ref{hermitian:partition}). At the same time 
  we can express the invariance of the integral under this symmetry in the form of constraints.
  To do this we collect the variation linear in $\epsilon_n$ under the integral (\ref{herm-matrix})
  which leads to the following expressions
  \bea
 && \langle  \sum\limits_{s=1}^{\infty}s\frac{t_s}{s!}\sum\limits_{i}\lambda_i^{s-1}\rangle~,\quad n=-1~,\label{expect:n-1}\\
 &&  \langle N^2 +  \sum\limits_{s=0}^{\infty}s\frac{t_s}{s!}\sum\limits_{i}\lambda_i^{s}\rangle~,\quad n=0~,\label{expect:n0}\\
   &&\langle(n+1) \sum\limits_i \lambda_i^n + \sum\limits_{i \neq j}
   \sum\limits_{k=0}^n \lambda_i^k  \lambda_j^{n-k} + \sum\limits_{s=0}^\infty s \frac{t_s}{s!} \sum_i \lambda_i^{s+n}\rangle~,
 \quad n\geq 1~,
  \eea
where expectation values are taken with respect to the partition function (\ref{herm-matrix}). 
Last expectation value can be rewritten combing the first and second terms leading to
    \bea
  \langle\sum\limits_{i, j} \sum\limits_{k=0}^n \lambda_i^k  \lambda_j^{n-k} + \sum\limits_{s=0}^\infty s \frac{t_s}{s!} \sum_i \lambda_i^{s+n}\rangle~,\quad n\geq1~.
  \label{expect:ng1}\eea
 Expressions (\ref{expect:n-1}),(\ref{expect:n0}) and (\ref{expect:ng1}) are 
 generated by the following operators acting  on $Z_N^{\rm herm}(\{t \}) $
  \bea
 && L_{-1}=\sum\limits_{k=1}^{\infty}t_k\frac{\partial}{\partial t_{k-1}}~,\quad
  L_0=\sum\limits_{k=0}^{\infty}k t_k\frac{\partial}{\partial t_{k}}+N^2\nonumber~,\\
  && L_n = \sum\limits_{k=0}^n (n-k)! k! \frac{\partial^2}{\partial t_k \partial t_{n-k}} +\sum\limits_{k=0}^\infty \frac{k (k+n)!}{k!} t_k \frac{\partial}{\partial t_{k+n}}
  ~,\quad n\geq1~.\label{vir-herm-matrix}
   \eea
 These are  the well-known  Virasoro constraints 
   \bea
    L_n  Z_N^{\rm herm}(\{t \}) =0~,~~~~n \geq -1~,
   \eea
  and  they satisfy the following algebra
   \bea
    [L_n, L_m ] = (n-m) L_{n+m}~.
   \eea
   
 Next let us consider the trigonometric version of the hermitian matrix model  
\bea
 Z^{\rm trig}_N(\{t_s\}) = \int \prod_{i} d\lambda_i~  \prod_{i\neq j} \sinh \big (\beta (\lambda_i - \lambda_j)\big ) ~e^{\sum\limits_{s=0}^\infty \frac{t_s}{s!} \sum\limits_{i} \lambda_i^s}~,\label{def-trig-mat}
\eea
 where we have introduced the deformation parameter $\beta$. 
The model is invariant under the following transformations
  \bea
   \lambda_i~\rightarrow~\lambda_i + \frac{ \epsilon_n}{2\beta} e^{2\beta n \lambda_i}~, ~~~n \geq -1~.
  \eea
We collect the  variation term linear in $\epsilon_n$ under the integral which leads to the following 
expectation values
\bea
&&\langle -N\sum_i e^{-2\beta \lambda_i}+\frac{1}{2\beta}\sum\limits_{s=1}^{\infty}\sum\limits_i s
\frac{t_s}{s!}e^{-2\beta \lambda_i}\lambda_i^{s-1}\rangle~,\quad n=-1~,\nonumber\\
&& \langle \frac{1}{2\beta}\sum\limits_{s=1}^{\infty}\sum\limits_{i}s\frac{t_s}{s!}\lambda_i^{s-1}\rangle~,\quad n=0~,\\
&& \langle  \sum\limits_{i, j}   \sum\limits_{k=0}^{n-1} e^{2\beta k \lambda_i} e^{2\beta (n-k) \lambda_j } 
 + \frac{1}{2\beta} \sum\limits_{s=1}^\infty s \frac{t_s}{s!} \sum\limits_i e^{2\beta n \lambda_i} \lambda_i^{s-1}\rangle
 ~,\quad  n\geq 1~.\nonumber
\eea
These terms are generated by the following operator
\bea
&&L_{-1}=-N\sum\limits_{k=0}^{\infty}(-2\beta)^l\frac{\partial}{\partial t_k}-\sum\limits_{k=1}^{\infty}\sum\limits_{l=0}^{\infty}
(-2\beta)^{l-1}t_k\frac{(l+k-1)!}{(k-1)!}\frac{\partial}{\partial t_{l+k-1}}~,\nonumber\\
&& L_0=\frac{1}{2\beta}\sum\limits_{k=1}^{\infty}t_k\frac{\partial}{\partial t_{k-1}}~,\\
&& L_n =    \sum\limits_{k=0}^{n-1} \sum\limits_{s=0}^\infty \sum\limits_{l=0}^\infty (2\beta k)^s (2\beta (n-k))^l \frac{\partial^2}{\partial t_s \partial t_l}
 +  \sum\limits_{k=1}^\infty   \sum\limits_{l=0}^\infty   (2\beta)^{l-1} n^l   \frac{(l+k-1)!}{(k-1)!~l!} t_s 
 \frac{\partial}{\partial t_{l+k-1}}~, \nonumber
\eea
which annihilates $Z^{\rm trig}_N(\{t_s\})$. 
 Using the binomial expansion we can show that these differential operators satisfy the Virasoro algebra
 \bea
  [L_n, L_m] = (n-m) L_{n+m}~.
 \eea
 Previously the Virasoro constraints for  $Z^{\rm trig}_N(\{t_s\})$ were discussed in \cite{Dubinkin:2013tda}, although in a bit less
  straightforward fashion.

\section{Toy model for $q$-Virasoro constraints}\label{sec-toy}

In this section we would like to reflect on the origin of Virasoro symmetry in integrals 
  and then generalize our observations to the case of $q$-Virasoro symmetry. To do this 
we will consider some toy examples of matrix models.
 On the way we will also introduce some basics in $q$-calculus and 
  necessary combinatorial tools. 

If we consider the functions in one variable $x$ then the classical 
Virasoro algebra has the following  well-known representation as first order differential operators
   \be
  L_n = -x^{n+1} \partial_x~.   
   \ee
   Alternatively there exists a different representation by the following operators
\be
 L_n = - (n+1) x^n - x^{n+1} \partial_x = - \partial_x (x^{n+1} ... )~.
\ee
 Consider the integral along the real line of a function $f(x)$
\be
 \int\limits^{\infty}_{-\infty} dx~ f(x)~,
\ee
 then,   provided the function $f(x)$ is differentiable and decays fast enough at infinity\footnote{  In order for 
 the integral (\ref{integral:f}) to be well defined function $f(x)$ should satisfy $\lim\limits_{|x|\to\infty}(x^n f(x))=0$
 for any positive $n$.}, this integral has the Virassoro symmetries
\be
 \int\limits^{\infty}_{-\infty} dx~ L_n f(x) = -\int\limits^{\infty}_{-\infty} dx~  \partial_x (x^{n+1}f(x))=0~.
 \label{integral:f}
\ee
 However these Virasoro symmetries cannot be converted to any PDEs since the integral is just a number.
  Instead we  can consider the generating function with infinitely many parameters
\be
 Z^{\rm toy} (\{t \})= \int dx~ x^\alpha e^{\sum\limits_{s=0}^\infty \frac{t_s}{s!} x^s}~,\label{toy-gener-func}
\ee
 which encodes many different integrals. 
 Then the condition 
 \be
\int dx~  L_n  \left  (  x^\alpha e^{\sum\limits_{s=0}^\infty \frac{t_s}{s!} x^s} \right ) =
- \int dx~ \partial_x \left  (x^{n+1}  x^\alpha e^{\sum\limits_{s=0}^\infty \frac{t_s}{s!} x^s} \right ) =0
 \ee
 implies the Virasoro constraints $L_n Z^{\rm toy} (\{t \})=0$ where $L_n$ is defined as the following differential 
  operator
  \be
  L_n =  (n+\alpha+1) n!\frac{\partial}{\partial t_n} +\sum\limits_{k=1}^\infty \frac{(k+n)!}{(k-1)!} t_k \frac{\partial}{\partial t_{k+n}} ~,~~~~
  n\geq 0~,\label{class-limit-Vir-toy}
 \ee
  which satisfy the Virasoro algebra.
   Thus for the case of the hermitian matrix model (\ref{herm-matrix})  the Virasoro operators (\ref{vir-herm-matrix}) can be derived 
    by inserting under the integral the following operators
\be
 L_n = - \sum\limits_i  \left ((n+1) \lambda_i^n + \lambda_i^{n+1} \partial_{\lambda_i} \right )
 = -  \sum\limits_i \partial_{\lambda_i}  (\lambda_i^{n+1} ... )~,
\ee
  which by themselves generate the Virasoro algebra.  
  The similar trick can be applied to the trigonometric generalization (\ref{def-trig-mat}). 

 Now let us introduce the basic notions of the $q$-calculus with $q$ being some complex number. 
  The quantum number $n$ is defined as 
 \bea
 [n]_q = \frac{q^n -1}{q-1}
 \eea
 and in the limits $q\rightarrow 1$ it becomes just $n$. 
 The quantum derivative is defined as the following difference operator
 \bea
  D_q f(x) = \frac{f(qx)-f(x)}{(q-1)x}~. \label{def-Dq}
 \eea
 Upon the limit $q\rightarrow 1$ the $q$-derivative $D_q$ becomes the ordinary derivative. 
 The $q$-derivative satisfies the modified Leibniz rule 
\bea
 D_q (f(x) g(x)) =g(qx) D_q f(x) + f(x) D_q g(x)\label{qD-Leibniz}
\eea
and we have 
\be
 D_q x^n = [n]_q x^{n-1}~.\label{qD-xn}
\ee
We can define the following   $q$-Virasoro operator
\bea
 T^q_n = -  D_q (x^{n+1} ...) ~,\label{def-q-Vir-oper}
\eea
 which acts on the function of one variable. 
 Alternatively we can define the operators $-x^{n+1} D_q$, but these two definitions lead 
 to the same algebraic properties. We will concentrate on 
  the definition (\ref{def-q-Vir-oper}) since it is the most suitable for the discussion of matrix models. 
 The operators (\ref{def-q-Vir-oper}) satisfy  the following relation 
\bea
 q^n T^q_n T^q_m - q^m T^q_m T^q_n = ([n]_q - [m]_q) T^q_{n+m}~, \label{qVir-lgebra}
\eea
or alternatively we can rewrite this as follows
\bea
 q^{n+1} T^q_n T^q_m - q^{m+1} T^q_m T^q_n = ([n+1]_q - [m+1]_q) T^q_{n+m}~. 
\eea
 In checking these relations we have to use the properties (\ref{qD-Leibniz}) and (\ref{qD-xn}).  
Equivalently the commutator of two generators (\ref{def-q-Vir-oper}) can be represented as
 \be
 [T^q_n, T^q_m] = - \sum\limits_{l=1}^{\infty} f_l (T^q_{n-l} T^q_{m+l} - T^q_{m-l} T^q_{n+l})~,
 \ee
  where  the coefficients\footnote{This form of deformed Virasoro algebra was introduced in \cite{Shiraishi:1995rp}
   with prescribed $f_l$. One can show that the realization (\ref{def-q-Vir-oper}) is a special case of their deformation
\cite{anton-fabrizio}.} $f_l$ are chosen such that $\sum\limits_{l=1}^\infty f_l q^l=1$. There is still  
   another relation we can write if we allow to introduce the generators depending on $q^2$
   \be
    [T_n^q, T_m^q] =  q^{-n-m}([n]_q - [m]_q) \left ( [2]_q T^{q^2}_{n+m} - T^q_{n+m} \right )~. \label{exotic-q-vir-alg}
   \ee 
   In principle we can go on and generate the whole tower of new operators $T^{q^i}_n$, $i =1,2,3...$ 
   and they will form infinite dimensional 
    Lie algebra.   We leave aside the details and other algebraic  properties of these generators for the future work
     \cite{anton-fabrizio}. 
    For us it is important 
     to remember  that the $q$-Virasoro generators (\ref{def-q-Vir-oper}) satisfy the 
     algebraic relation (\ref{exotic-q-vir-alg}). 
  
   The crucial property of $D_q$ is that its integral over the line (even over the half-line $[0,\infty)$) 
    is identically  zero
   \be
    \int\limits_{-\infty}^\infty dx D_q f(x) =  \frac{1}{q-1} \int\limits_{-\infty}^\infty dx \frac{f(qx)}{x} -  
    \frac{1}{q-1} \int\limits_{-\infty}^\infty dx \frac{f(x)}{x} =0~. 
   \ee
    Thus an integral over the line vanishes upon the insertion of 
    the operators $D_q (x^{n+1} ...)$ under the integral. 
    Hence if we take the  toy generating function (\ref{toy-gener-func})
  and insert the operators $D_q (x^{n+1} ...)$  we get the following set of identities
\be
 \int dx~ D_q \left ( x^{n+1} x^\alpha~ e^{\sum\limits_{s=0}^\infty \frac{t_s}{s!} x^s}\right )=0~.
\ee
   The exponent can be expanded as follows
   \be
    e^{\sum\limits_{s=1}^\infty \frac{t_s}{s!} x^s} = \sum\limits_{n=0}^\infty B_n (t_1, t_2, ... , t_n) \frac{x^n}{n!}~,\label{Bell-exponent}
   \ee
where   $B_n$ is  the $n$th complete Bell polynomial defined as
   \be
    B_n(t_1, t_2, ... , t_n) ={\rm det }\left|\begin{array}{cccccccc}
               t_1 & \binom{n-1}{1}  t_2 & \binom{n-1}{2} t_3 & \binom{n-1}{3} t_4 & \binom{n-1}{4} t_5 & ... & ... & t_n\\
               -1 & t_1 & \binom{n-2}{1} t_2 & \binom{n-2}{2} t_3 & \binom{n-2}{3} t_4 & ... & ... & t_{n-1} \\
               0 & -1 & t_1  & \binom{n-3}{1} t_2 & \binom{n-3}{2} t_3 & ... &... & t_{n-2} \\
               0 & 0 & -1 & t_1 & \binom{n-4}{1} t_2 & ... & ... & t_{n-3} \\
               0 & 0 & 0& -1 & t_1 & ... & ... & t_{n-4} \\
               0 & 0 & 0 & 0& -1 & ... & ... & t_{n-5}\\
               ... & ... & ... & ...&  ... & ... & ... & ...\\
               0 & 0 & 0 & 0& 0 & ... & -1 & t_1\\
        \end{array} \right |~,
   \ee
    where $\binom{n}{k}$ stands for the binomial coefficient $\frac{n!}{k! (n-k)!}$. 
    The complete Bell polynomials are weighted homogeneous polynomials. For instance, 
    the first few Bell polynomials are:
     \be
 &&    B_1(t_1) =t_1~,\nonumber\\
  &&   B_2(t_1, t_2) = t_1^2 + t_2~,\nonumber \\
   && B_3 (t_1, t_2, t_3) = t_1^3 + 3 t_1 t_2 + t_3~,\\
   && B_4 (t_1, t_2, t_3, t_4) = t_1^4 + 6 t_1^2 t_2 + 4 t_1 t_3 + 3 t_2^2 + t_4~,\nonumber  
     \ee
   and we assume that $B_0=1$.   
   We will use the following properties of the Bell polynomials
       \be
   B_l \left ( (\alpha + \beta) t_1, ... , (\alpha + \beta) t_l \right ) = \sum\limits_{p=0}^l \binom{l}{p} B_{l-p} \left ( \alpha  t_1, ... , \alpha  t_{l-p} \right ) 
    B_p \left ( \beta t_1, ... , \beta t_p \right )~,\label{Bell-pr1}
  \ee
 and 
     \be
    \frac{\partial}{\partial t_l} B_n (t_1, ... , t_n) = \left \{ \begin{array}{ll}      
                                                                              0~, & n<l\\
                                                                               \frac{n!}{(n-l)! l!} ~B_{n-l} (t_1, ... , t_{n-l})~, & n \geq l
                                                                               \end{array} \right .  ~,\label{Bell-pr2}
     \ee
     and
     \be
      B_l (\tilde{t}_1, ... ,\tilde{t}_l) = \sum\limits_{p=0}^l q^p \binom{l}{p} B_p (t_1, ... , t_p) B_{n-p} (-t_1, ... , - t_{n-p})~,
      \label{Bell-pr3}
     \ee
      where $\tilde{t}_k = (q^k-1) t_k$. 
      These properties are easily derivable from the definition (\ref{Bell-exponent}). 
   
     After applying the definition (\ref{def-Dq}) and recombining the terms we arrive to the constraints
   \be
    T^q_n Z^{\rm toy} (\{t \})=0~,~~~~~n\geq 0~,
   \ee
    where the operators $T_n^q$ are defined as follows
    \be
     T^q_n = \frac{1}{q-1}\left[\sum\limits_{l=0}^\infty \sum\limits_{k=0}^l \frac{[n+l-k+\alpha +1]_q}{(l-k)! k!} B_{l-k} (t_1, t_2, ...., t_{l-k}) B_k (-t_1, -t_2, ... , -t_k) \frac{\partial}{\partial t_{n+l}}
    \right.\nonumber\\\left.
-n!\frac{\partial}{\partial t_n} \right]   ~.\label{toy-Tn-def}
   \ee
   Alternatively  using the property (\ref{Bell-pr3}) we can rewrite the operator (\ref{toy-Tn-def}) as 
   \be
    T_n^q =   [n+\alpha +1]_q  n!\frac{\partial}{\partial t_n} + \frac{q^{n+\alpha +1}}{q-1} \sum\limits_{k=1}^\infty
    \frac{(k+n)!}{k!} B_k (\tilde{t}_1, ... , \tilde{t}_k) \frac{\partial}{\partial t_{k+n}}~,\label{toy-Tn-def2}
   \ee
   with $\tilde{t}_k = (q^k-1) t_k$.
    One can observe that the operator (\ref{toy-Tn-def2}) collapses to the operator (\ref{class-limit-Vir-toy})
     in the classical limit $q \rightarrow 1$. We have checked explicitly that the operators $T^q_n$
      satisfy the algebra (\ref{exotic-q-vir-alg}).  In checking the algebra (\ref{exotic-q-vir-alg}) we had to 
       use the properties (\ref{Bell-pr1}) and (\ref{Bell-pr2}).

\section{$q$-Virasoro   for elliptic hermitian matrix model}\label{sec-q-Vir}

In this section we derive the $q$-Virasoro constraints for the elliptic generalisation of the matrix model. 
 Although any matrix integral will vanish under the insertion of $D_q$ 
 only in the case of the elliptic matrix model the operator $D_q$ talks nice to the matrix model measure.  

The partition function for the $4d$ $\cal{N}$$= 1$ $U(N)$ gauge theory on $S^3 \times S^1$
 \be
  \int \prod_{i=1}^N
   d\lambda_i \prod_{i<j}\left(1- e^{i(\lambda_i-\lambda_j)}\right)\left(1- e^{i(\lambda_j-\lambda_i)}\right)
 \prod_{n=1}^{\infty}\left(1-q^n e^{i(\lambda_i-\lambda_j)}\right)^2\left(1-q^{n} e^{i(\lambda_j-\lambda_i)}\right)^2~,
 \ee
 where the integration is over the real line and $q\equiv e^{\beta}$ with $\beta$ being the circumference of $S^1$.
 Performing the change of variables $e^{i\lambda_i}=z_i$ we arrive at the following form of the partition function 
 
 \be
    \oint \prod_{i=1}^N \frac{d z_i}{z_i}   \prod_{i<j}\theta\left(\frac{z_i}{z_j};q\right)
  \theta\left(\frac{z_j}{z_i};q\right)~,
 \ee
 where the integration is now over the contour around the origin and $\theta$ function is defined as in (\ref{def-theta}). 
Next, we can introduce the generating function
  \be 
 Z_N^{\rm ell}(\{ t\})= \oint \prod_{i=1}^N \frac{d z_i}{z_i}   \prod_{i<j}\theta\left(\frac{z_i}{z_j};q\right)
  \theta\left(\frac{z_j}{z_i};q\right) e^{\sum\limits_{k=0}^{\infty} \frac{t_k}{k!}\sum\limits_{i=1}^N z_i^k  }~,
  \label{wilson:loop}
 \ee
 for the expectation values of supersymmetric Wilson loop in different representations. 
 Our goal is to show that $ Z_N^{\rm ell}(\{ t\})$ is annihilated by the $q$-Virasoro constraints. 
 
Following the logic from the previous section we can define the differential operator 
  \be
   T^q_n = -\sum\limits_{l=1}^N D_q^{z_l}(z_l^{n+1}\dots)~,\label{qVir-matr-op}
   \ee
   which acts on the functions of $N$ variables  and $ D_q^{z_l}$ is $q$-derivative with respect to the $z_l$-variable.
   These operators satisfy the  $q$-Virasoro algebra (\ref{qVir-lgebra})-(\ref{exotic-q-vir-alg}). 
    The insertion of these operators under the contour integral (\ref{wilson:loop}) gives us identically zero. 
    Now we have to analyse in details how these differential operators act on the integrand. 
     Using the properties of $\theta$ function 
    \be
     \theta(qz; q) = \theta (z^{-1}; q)~,~~~~ \theta(q^{-1}z; q) =  q^{-1} z^2~ \theta (z^{-1}; q)~,
    \ee
   we arrive at the following relation  
 \be
 &&\sum_{l=1}^N D_q^{z_l}\left(f(z_l)\prod_{i<j}\theta\left(\frac{z_i}{z_j};q\right)\theta\left(\frac{z_j}{z_i};q\right)\right)= \label{useful-id}\\
 &&\sum_{l=1}^N\frac{1}{(q-1)z_l}\left(\frac{f(qz_l)}{f(z_l)}\prod_{j\neq l}q^{-1}\frac{z_j^2}{z_l^2}-1\right)
 f(z_l)\prod_{i<j}\theta\left(\frac{z_i}{z_j};q\right)\theta\left(\frac{z_j}{z_i};q\right)~.  \nonumber
 \ee
For  the calculation of the $q$-derivative of the exponental factor  we use the expansion (\ref{Bell-exponent})  in terms of the  Bell 
polynomials $B_k$
\be\nonumber
&&\exp\left(\sum\limits_{k=1}^{\infty}\frac{t_k}{k!}q^k z_i^k\right)=\sum_{k=0}^{\infty}\frac{1}{k!}B_k(t_1,\dots, t_k)
q^kz_i^k=\\
&&\sum_{k=0}^{\infty}\sum_{p=0}^{\infty}\frac{1}{k!p!}B_k(t_1,\dots, t_k)B_p(-t_1,\dots, -t_p)q^k z_i^{k+p}
\exp\left(\sum\limits_{l=1}^{\infty}\frac{t_l}{l!} z_i^l\right)=\nonumber\\
&&\sum_{k=0}^{\infty}\frac{1}{k!}B_k\left(\tilde{t}_1,\dots,\tilde{t}_k\right)x^k\exp\left(\sum\limits_{l=1}^{\infty}\frac{t_l}{l!} z_i^l\right)
.\label{Bells-useful}
\ee
Applying the formulas (\ref{useful-id}) and (\ref{Bells-useful}) we find that the insertion of the operator $T^q_n$ (\ref{qVir-matr-op}) under   the integral (\ref{wilson:loop}) is equivalent to the insertion of the following
 terms under the integral
\be
  \frac{1}{q-1}\left[\prod_{j=1}^{N} z_j^2 \sum_{l=1}^{N}\sum_{k,p=0}^{\infty}q^{n+1+k-N}\frac{1}{k!p!}B_k(t_1,\dots,t_k)
B_{p}(-t_1,\dots,-t_{p})z_l^{k+p+n-2N} - \sum_{l=1}^{N} z_l^n\right]~.\label{gen-terms-elvir}
\ee
 Thus the expectation value of these terms should be zero. 
  Now our final goal is to  generate these terms by taking the appropriate 
  $t$-derivatives of the integrand of  (\ref{wilson:loop}). In order to do it
 we need to rewrite $\prod\limits_{i=1}^N z_i$ in terms of sums  $\sum\limits_{i=1}^N z_i^k$.
 This can be done using the Newton's identities 
 \be
 \prod_{i=1}^N z_i=\frac{1}{N!}\left|\begin{array}{cccccc}
  p_1 & 1 & 0 & \dots &  &  \\
  p_2  & p_1 & 2 & 0 & \dots & \\
  \dots   &  \dots  &\dots &  \dots  &\dots &\\
  p_{N-1} & p_{N-2} & \dots  &\dots& p_1 & N-1\\
   p_{N} & p_{N-1} & \dots  &\dots& p_2 & p_1 
 \end{array}\right|~,
 \ee
 where $p_k\equiv \sum\limits_{i=1}^{N}z_{i}^k $.  The terms $\sum\limits_{i=1}^{N}z_{i}^k$ can be generated by taking 
  the $t$-derivatives  and thus we can introduce the following differential operator
   \be
 {\cal D}_N =\frac{1}{N!}\left|\begin{array}{cccccc}
 2! \frac{\partial}{\partial t_2} & 1 & 0 & \dots &  &  \\
  4!\frac{\partial}{\partial t_4}  &2! \frac{\partial}{\partial t_2} & 2 & 0 & \dots & \\
  \dots   &  \dots  &\dots &  \dots  &\dots &\\
 (2N-2)! \frac{\partial}{\partial t_{2N-2}} & (2N-4)!\frac{\partial}{\partial t_{2N-4}} & \dots  &\dots& 2!\frac{\partial}{\partial t_2} & N-1\\
  (2N)! \frac{\partial}{\partial t_{2N}} & (2N-2)!\frac{\partial }{\partial t_{2N-2}} & \dots  &\dots& 4!\frac{\partial}{\partial t_4} & 2! \frac{\partial}{\partial t_2} 
 \end{array}\right|~,
 \ee
 with the property that
 \be
  \prod_{j=1}^{N} z_j^2 ~ e^{\sum\limits_{k=0}^{\infty} \frac{t_k}{k!}\sum\limits_{i=1}^N z_i^k  }=
   {\cal D}_N  \left (e^{\sum\limits_{k=0}^{\infty} \frac{t_k}{k!}\sum\limits_{i=1}^N z_i^k  } \right )~.
 \ee
 Combining all together we obtain the following $q$-Virasoro operator
 \be
 T^q_n = \frac{1}{q-1}\left[  \sum_{k,p=0}^{\infty}q^{n+1+k-N}\frac{(k+p+n-2N)!}{k!p!}B_k(t_1,\dots,t_k)
B_{p}(-t_1,\dots,-t_{p})\times \right.\\\left. {\cal D}_N \frac{\partial}{\partial t_{k+p+n-2N}} - 
 n!\frac{\partial}{\partial t_n}\right]~,\label{qVir-exp-matrix}
 \ee
 which annihilates the generating function $Z_N^{\rm ell}(\{ t\})$. Using the property (\ref{Bell-pr3}) we
  can rewrite the operator (\ref{qVir-exp-matrix}) as follows
  \be
   T_n^q=\frac{1}{q-1}\left[ q^{n+1-N} \sum\limits_{l=0}^\infty \frac{(l+n-2N)!}{l!} B_l(\tilde{t}_1, ... , \tilde{t}_l)
   {\cal D}_N \frac{\partial}{\partial t_{l+n-2N}} - n!\frac{\partial}{\partial t_n}\right]~.
   \label{qVir:operator}
  \ee
  We see that generically the operators $T_n^q$ are higher order differential operators and action of these operators
   on $Z_N^{\rm ell}(\{ t\})$ generates the insertion of the terms (\ref{gen-terms-elvir}) under the integral. 
    It is crucial to stress that there are many different higher order operators which will generate exactly the same 
     insertion  (\ref{gen-terms-elvir}). 
   This fact
  complicates the calculation of the algebra and we elaborate more on this point later. 
    In the next section we will show that the operators (\ref{qVir:operator}) satisfy the following algebra
   \be
    [T_n^q, T_m^q] = q^{-n-m}([n]_q - [m]_q) \left ( [2]_q T^{q^2}_{n+m} - T^q_{n+m} \right )~,
\label{qVir:new}    
   \ee 
 where the operators $T^{q^2}_n$ are defined below in (\ref{vir:squared}). The operators $T^{q^2}_n$ annihilate 
  $Z_N^{\rm ell}(\{ t\})$. Indeed if we continue to calculate the algebra we will get the whole tower of operators
   $T^{q^j}_n$, $j=1,2, 3, ...$ which annihilate  $Z_N^{\rm ell}(\{ t\})$. 
  
  In order to define explicitly the operators $T^{q^2}_n$ we have to insert under the integral (\ref{wilson:loop}) the
  following difference operator
  \be
   T^{q^2}_n = -\sum\limits_{l=1}^N D_{q^2}^{z_l}(z_l^{n+1}\dots)~.
   \ee
  Using the following properties of the $\theta$ function
\be
\theta\left(q^2 z;q\right)=-q^{-1}z^{-1}\theta\left(z^{-1};q\right)~,\qquad
\theta\left(q^{-2} z;q\right)=-q^{-3}z^{3}\theta\left(z^{-1};q\right)~,
\ee
we obtain following Ward identities 
 \be
 \left\langle\frac{1}{q^2-1}\left[\prod_{j=1}^{N} z_j^4 \sum_{l=1}^{N}\sum_{p=0}^{\infty}
 q^{2n+4-4N}\frac{1}{p!}B_p(\hat{t}_1,\dots,\hat{t}_p)
z_l^{p+n-4N} - \sum_{l=1}^{N} z_l^n\right]\right\rangle=0~,
 \ee
  where by $\langle ... \rangle$ we mean the insertion of this expression under the integral (\ref{wilson:loop}). 
 These Ward identities can be expressed in the form $T_{n}^{q^2} Z_N^{\rm ell}(\{ t\})=0$, where the differential 
 operator $T_{n}^{q^2}$ is given by 
 \be
 T_{n}^{q^2}=\frac{1}{q^2-1}\left[ q^{2n+4-4N} \sum\limits_{l=0}^\infty \frac{(l+n-4N)!}{l!} B_l(\hat{t}_1, ... , \hat{t}_l)
   {\cal \tilde{D}}_N \frac{\partial}{\partial t_{l+n-4N}} - n!\frac{\partial}{\partial t_n}\right]~.
   \label{vir:squared}
 \ee

 Here ${\cal \tilde{D}}_N$ is the differential operator defined as 
  \be
 {\cal \tilde{D}}_N =\frac{1}{N!}\left|\begin{array}{cccccc}
 4! \frac{\partial}{\partial t_4} & 1 & 0 & \dots &  &  \\
  8!\frac{\partial}{\partial t_8}  &4! \frac{\partial}{\partial t_4} & 2 & 0 & \dots & \\
  \dots   &  \dots  &\dots &  \dots  &\dots &\\
 (4N-4)! \frac{\partial}{\partial t_{4N-4}} & (4N-8)!\frac{\partial}{\partial t_{4N-8}} & \dots  &\dots& 4!\frac{\partial}{\partial t_4} & N-1\\
  (4N)! \frac{\partial}{\partial t_{4N}} & (4N-4)!\frac{\partial }{\partial t_{4N-4}} & \dots  &\dots& 8!\frac{\partial}{\partial t_8} & 4! \frac{\partial}{\partial t_4} 
 \end{array}\right|~.
 \ee
 Again we see that the operators $T_{n}^{q^2}$ are higher order differential operators. Analogously we can define 
  the operators $T^{q^j}_n$.

 In the expression (\ref{qVir-exp-matrix})  there is a problem  since the operator $T^q_n$ is defined only for the case $n\geq 2N$. 
  Similar problems exist for the operator $T^{q^2}_n$ which is defined for the case  $n\geq 4N$. 
 To resolve this problem we can insert the different $q$-Virasoro operator $ -\sum\limits_{l=1}^N D_q^{z_l}(z_l^{2N(n+1)+1}\dots)$
 under the integral. This leads to the following differential operator
 \be
\tilde{T}^q_n=  \sum_{p\geq0}^{\infty}q^{2N(n+1)+1-N}\frac{(p+2Nn)!}{p!}B_k(\tilde{t})
{\cal D}_N \frac{\partial }{\partial t_{p+2Nn}} - (2Nn+2N)!\frac{\partial}{\partial t_{2N(n+1)}}~, \label{qVir-mod}
\ee
  so that the operator is well defined for any $n\geq -1$. The algebra can be calculated in completely analogous way. 
 
\section{Calculation of the algebra}\label{s-calculation}

 In this section we derive the algebra (\ref{qVir:new}) for the operators $T^q_n$ defined in (\ref{qVir:operator}) and 
  the operators $T^{q^2}_n$ defined in (\ref{vir:squared}). The operators $T^q_n$ and $T^{q^2}_n$ are higher order 
   differential operators which generate concrete insertions under the integral. However 
    different differential operators can generate exactly the same insertions under the integral. Thus in principle 
     we can have different representations for  $T^q_n$ and $T^{q^2}_n$ as higher order differential operators in $t$'s. 
     We have to keep in   mind this feature of these operators. 
    
Our goal is to check the relation (\ref{qVir:new})  
 for the operators (\ref{qVir:operator}). For this we calculate the commutator $[T_n^q, T_m^q]$. 
  In order to calculate it we need to know the action of the operator 
   ${\cal D}_N$  on the complete Bell polynomials. One can easily derive the following relation
   \be
    {\cal D}_N \left (e^{\sum\limits_{s=1}^\infty \frac{t_s}{s!} x^s} \right ) =0~,
   \ee
  which upon the expansion (\ref{Bell-exponent}) implies that the operator ${\cal D}_N$ annihilates the Bell polynomials
 \be
 {\cal D}_N B_{k}(t_1,\dots,t_k) =0~.\label{newton:bell1}
 \ee
  Next we calculate the following relation
   \be
    {\cal D}_N \left (e^{\sum\limits_{s=1}^\infty \frac{\tilde{t}_s}{s!} x^s} \right ) =  (-1)^{N+1} (q^2-1) x^{2N} e^{\sum\limits_{s=1}^\infty \frac{\tilde{t}_s}{s!} x^s}  ~,
   \ee
   where we recall that $\tilde{t}_s = (q^s-1)t_s$. Expanding this formula in $x$ we get the following action of ${\cal D}_N$ 
 \be
 \label{newton:bell2}{\cal D}_N B_{k}(\tilde{t}_1,\dots,\tilde{t}_k)&=&\left \{ \begin{array}{ll}      
                                                                              0~, & k<2N\\
                                                                               (-1)^{N+1}(q^2-1)\frac{k!}{(k-2N)!}B_{k-2N}(\tilde{t}_1,\dots,\tilde{t}_{k-2N})~, & k\geq 2N                                                                             
                                                                               \end{array} \right . 
\ee                                                                               
   Finally we need the following relation  
 \be                                                                              
 \label{bell:relation:hat}B_p(\hat{t}_1,\dots,\hat{t}_p)&=&\sum\limits_{k\leq p}\frac{p!}{k!(p-k)!}q^k B_k(\tilde{t}_1,\dots,\tilde{t}_k)B_p(\tilde{t}_1,\dots,\tilde{t}_p)\,,
 \ee
 where $\hat{t}_k\equiv (q^k+1)\tilde{t}_k=(q^{2k}-1)t_k$. This relation follows directly from the expansion (\ref{Bell-exponent}).
  Using these formulas we can straightforwardly calculate $[T_n^q, T_m^q]$. 
 However for the case of arbitrary number $N$ of eigenvalues this problem is complicated and the corresponding expressions 
  are enormous. Thus we will  concentrate on the simplest cases of $N=2$ and $N=3$.
  
 Using the above formulas it is straightforward to evaluate 
 commutator of $T_n^q$ operators for the case $N=2$  
\be
&&\left[T_n^q,T_m^q\right] =\nonumber \\
&&\frac{q^{-n-m}}{(q-1)^2}\left([n]_q-[m]_q\right)\left[q^{2n+2m-6}
\sum\limits_{p\geq 0}\frac{1}{p!}B_p(\hat{t}){\cal D}_2\left({\cal D}_2 (p+n+m-8)!\frac{\partial}{\partial t_{p+n+m-8}}
-\right. \right.\nonumber \\
&&\nonumber\left.
 (q^2-1)(p+n+m-4)!\frac{\partial}{\partial t_{p+n+m-4}}
+(q^2-1)2!(p+n+m-6)!\frac{\partial^2}{\partial t_2 \partial t_{p+n+m-6}}\right)-\\
&&\left. q^{n+m-1}\sum\limits_{p\geq 0}\frac{(p+n+m-4)!}{p!}B_p(\tilde{t}){\cal D}_2\frac{\partial}{\partial t_{p+n+m-4}}\right]~.
\label{vir:commutator:2}
\ee

For the case of $N=3$ we obtain

\be
&&\left[T_n^q,T_m^q\right]= \nonumber \\
&&\frac{q^{-n-m}}{(q-1)^2}\left([n]_q-[m]_q\right)\left[q^{2n+2m-10}
\sum\limits_{p\geq 0}\frac{1}{p!}B_p(\hat{t}){\cal D}_3\left({\cal D}_3 (p+n+m-12)!\frac{\partial}{\partial t_{p+n+m-12}}
+\right. \right.\nonumber \\
&&\nonumber
(q^2-1)(p+n+m-6)!\frac{\partial}{\partial t_{p+n+m-6}} -\\
&& \frac{1}{2}(q^2-1)\left(4!\frac{\partial}{\partial t_4}-
(2!)^2\frac{\partial^2}{\partial t_2^2}\right)(p+n+m-10)!\frac{\partial}{\partial t_{p+n+m-10}}-\nonumber\\
&&\left. (q^2-1)2!(p+n+m-8)!\frac{\partial^2}{\partial t_2\partial t_{p+n+m-8}}\right)
  - \nonumber\\
  && \left. q^{n+m-2}\sum\limits_{p\geq 0}\frac{(p+n+m-6)!}{p!}B_p(\tilde{t}){\cal D}_3\frac{\partial}{\partial t_{p+n+m-6}}\right]~.
  \label{vir:commutator:3}
\ee

The last terms in both (\ref{vir:commutator:2}) and (\ref{vir:commutator:3}) contributes to $T^{q}_{n+m}$ 
operator in (\ref{qVir:new}). However to completely match 
these expressions with the commutation relations (\ref{qVir:new}) we need the explicit 
form (\ref{vir:squared}) of the operator $T^{q^2}_n$.  Notice that  ${\cal \tilde{D}}_N\sim\left({\cal {D}}_N\right)^2$, where $\sim$ means that these two operators are equivalent 
 upon the action on the partition function (\ref{wilson:loop}). As we can see the expressions in (\ref{vir:commutator:2}) and (\ref{vir:commutator:3})
 are complicated and do not match the operator (\ref{vir:squared}). However if we act with  
 the right hand side  of (\ref{vir:commutator:2})
 on the partition function we obtain the familiar terms. This action can be obtained directly by the substitution 
 \be
 \frac{\partial}{\partial t_a}\to (x_1^a+x_2^a)~, \quad {\cal D} \to x_1^2x_2^2~.
 \ee
 Then the operators written in the first two lines of (\ref{vir:commutator:2}) result in the following expectation value:
 \be
 &&\frac{([n]_q -[m]_q)}{(q-1)}\langle q^{n+m-6}\sum\limits_{p\geq 0}\frac{1}{p!}B_p(\hat{t})x_1^2x_2^2\left(
 (x_1^{a-8}+x_2^{a-8})x_1^2x_2^2-(q^2-1)(x_1^{a-4}+x_2^{a-4})+\right.\nonumber\\
&& \left.(q^2-1)(x_1^{a-6}+x_2^{a-6})(x_1^2+x_2^2)\right)\rangle= \frac{([n]_q - [m]_q)}{(q-1)}\langle
 q^{n+m-4}\sum\limits_{p\geq 0}\frac{1}{p!}B_p(\hat{t})x_1^4x_2^4(x_1^{a-8}+x_2^{a-8})\rangle\nonumber=\\
 &&([n]_q -[m]_q)\left[[2]_qT_{n+m}^{q^2}+n!\frac{\partial}{\partial t_n}\right]Z~,
 \ee
 where for shortness we have introduced $a=p+n+m$. Combining these terms with the last term in (\ref{vir:commutator:2})
 \be
 q^{n+m-1}\sum\limits_{p\geq 0}\frac{(p+n+m-4)!}{p!}B_p(\tilde{t}){\cal D}_2\frac{\partial}{\partial t_{p+n+m-4}}=
 T_{n+m}^q+n!\frac{\partial}{\partial t_n}
 \ee
 we can arrive to the desired commutation relation (\ref{qVir:new}). One can perform 
 similar calculation for the case $N=3$ by making the substitution 
 
  \be
 \frac{\partial}{\partial t_a}\to (x_1^a+x_2^a+x_3^a)~, \quad {\cal D} \to x_1^2x_2^2x_3^2~.
 \ee
 After some simple algebra one can show that the commutation relation (\ref{vir:commutator:3}) is compatible 
 with (\ref{qVir:new}) once we act on the partition function (\ref{wilson:loop}). 
 
 We have obtained the desired algebra  (\ref{qVir:new}) but on the way we had to perform some additional 
  manipulations. Let us provide the general explanation for what we did. For this purpose we consider general 
  matrix model of the form
 \be
  Z(\{t\}) = \int d^N x ~f(x_1, ..., x_N)~ e^{\sum\limits_{k=0}^\infty \frac{t_k}{k!} \sum\limits_{i=1}^N x_i^k}~,
 \ee
  where the $t$'s are parameters (either finite or infinite number of them). We are interested to discuss the symmetries of 
   this integral, namely the differential operators $D$ in $t$'s which annihilate $Z(\{t\})$
   \be
    D Z(\{t\}) =  \int d^N x ~f(x_1, ..., x_N)~\sigma_D(x_1, ..., x_N) ~e^{\sum\limits_{k=0}^\infty \frac{t_k}{k!} \sum\limits_{i=1}^N x_i^k}=0~.
   \ee
   The operator $D$ upon acting on the exponent generates the function $\sigma_D$. 
   However there is no unique correspondence between $D$ and the function $\sigma_D$.  
   Two different operators $D$ and $\tilde{D}$ generate the same function $\sigma_D$ if
     \be
      (D - \tilde{D})~ e^{\sum\limits_{k=0}^\infty \frac{t_k}{k!} \sum\limits_{i=1}^N x_i^k} =0~.\label{vir-ideal}
     \ee
 The operators which annihilate the exponent form an ideal among the differential operators. When we study the algebra 
  of the symmetries we have to quotient by this ideal. 
   For the symmetries $D_1$ and $D_2$
     we have 
    \be
     [D_1, D_2] Z(\{ t\})=0~.
    \ee
    All the symmetries will form a Lie algebra which  is the Lie algebra of the operators annihilating $Z$ modulo 
     the ideal discussed above. In calculation of the algebra (\ref{qVir:new}) we had to use some identification 
     (\ref{vir-ideal}). 
  Notice that this is a generic feature of the matrix models and even the simple Virasoro operator (\ref{vir-herm-matrix}) may have 
   a different representations upon these identifications.

\section{Summary}\label{sec-end}

The Virasoro constrains are important for the understanding of the hermitian matrix model. 
In this work we looked at the elliptic generalization of the hermitian matrix model and we
have derived the $q$-Virasoro constraints for this model. 
 Our deformation of the Virasoro algebra is based on the realization in terms of $q$-derivatives
 within the $q$-calculus 
 and this deformation is a special case 
  of a more general elliptic deformation of  the Virasoro algebra.  
 The deformation of Virasoro algebras has been  first discussed by  Curtright and  Zachos in \cite{Curtright:1989sw} (see
  also \cite{Sato:1991sj} for further explanation and the relevant references). 
  Later the deformation of the Virasoro algebra was introduced in \cite{Shiraishi:1995rp,Frenkel,Lukyanov:1994re} in a different context. 
   There were numerous works on the study of these deformations including different physical realizations of 
   $q$-Virasoro algebra (\cite{Nieri:2013yra,Nieri:2013vba,Awata:2010yy,Aganagic:2013tta,Morozov:2015xya,Zenkevich:2014lca,
   Mironov:2011dk,Itoyama:2013mca,Itoyama:2014pca}). However the realization of the general elliptic deformation 
  of Virasoro algebra  in terms of concrete difference operators is still missing. Moreover, many algebraic aspects remain 
   a mystery.  For example, the realization of the algebra in (\ref{exotic-q-vir-alg}) was forced upon us by the matrix model and 
    came as total surprise. 
 
 The trigonometric and elliptic deformations of the Virasoro algebra play a crucial role in the higher 
 dimensional gauge theories, as an
  example see the recent papers  \cite{fabrizio-new} and \cite{new-elliptic} on the role of the 
  elliptic deformation. What we have observed 
   in this paper is just a tip of the iceberg.  We think that the  $q$-Virasoro constraints are 
  a generic feature of the partition functions on $S^3 \times S^1$ for different gauge theories. 
  We believe that  the elliptic 
   deformation of the Virasoro algebra  should appear when one tries to generalise the present analysis to a wider
   class of theories. 
    However in order to make a further progress we need to understand better the algebraic property of
    the deformed Virasoro algebra  
     and to find the realization in terms of the concrete difference operators. We plan to answer these 
     questions in a forthcoming work
      \cite{anton-fabrizio}.

\bigskip
{\bf Acknowledgements}  We thank Reimundo Heluani and Fabrizio Nieri for the discussions.  
 We are grateful to Fabrizio Nieri for the reading and the commenting on the manuscript.
  M.Z. is grateful to IMPA, Rio de Janeiro for the hospitality where the work has been finished.
The research of M.Z. is supported in part by Vetenskapsr\r{a}det 
 under grant \#2014- 5517, by the STINT grant and by the grant  Geometry and Physics 
  from the Knut and Alice Wallenberg foundation. The research of A.N. is supported in part by 
  INFN and by MIUR-FIRB grant RBFR10QS5J "String Theory and Fundamental Interactions".    


  \begingroup\raggedright\endgroup

\end{document}